\documentclass{PoS}

\usepackage{wrapfig}

\newcommand\beq{\begin{eqnarray}}
\newcommand\eeq{\end{eqnarray}}
\newcommand {\bmp}{\begin{minipage}}
\newcommand {\emp}{\end{minipage}}
\newcommand \reffig {Fig.~\ref}

\title{Lattice calculation for unitary fermions in a finite box}

\ShortTitle{Lattice calculation for unitary fermions in a finite box}

\author{\speaker{Jong-Wan Lee}\\
        Institute for Nuclear Theory, University of Washington, Seattle, WA 98195-1550\\
        E-mail: \email{jwlee823@washington.edu}}

\author{Michael G. Endres\\
		Physics Department, Columbia University, New York, NY 10027, USA and\\
		Theoretical Physics Laboratory, RIKEN, Wako, Saitama 351-0198, Japan\\
		E-mail: \email{endres@riken.jp}}

\author{David B. Kaplan\\
		Institute for Nuclear Theory, University of Washington, Seattle, WA 98195-1550\\
		E-mail: \email{dbkaplan@washington.edu}}

\author{Amy N. Nicholson\\
		Institute for Nuclear Theory, University of Washington, Seattle, WA 98195-1550\\
		E-mail: \email{amynn@washington.edu}}

\abstract{A fundamental constant in systems of unitary fermions is the so-called Bertsch parameter, the ratio of the ground state energy for spin paired unitary fermions to that for free fermions at the same density. I discuss how we computed this parameter as well as the pairing gap using a recently developed lattice construction for unitary fermions, by measuring correlation functions for up to 38 fermions in a finite box. Our calculation illustrates interesting issues facing the study of many-body states on the lattice, which may eventually be confronted in QCD calculations as well.}

\FullConference{The XXVIII International Symposium on Lattice Field Theory, Lattice2010\\
		June 14-19, 2010\\
		Villasimius, Italy}

\begin{document}

\section{Introduction}
Fermions at unitarity is an idealized system which has an infinite $S$-wave scattering length and zero interaction range or equivalently the two-particle s-wave scattering phase shift $\delta_0=\pi/2$. In the unitarity limit, the only relevant scale is the density ($\rho$) and the ground state energy is related to that of the non-interacting fermions by
\beq
E_{unitary}(\rho)=\xi E_{free}(\rho),
\eeq
where the dimensionless parameter $\xi$ is called the Bertsch parameter. $\xi$ is of our particular interest because it is the unique parameter which relates zero temperature thermodynamic quantities between unitary and free Fermi gas. Another important physical quantity is the pairing gap ($\Delta$), the difference between the chemical potential and the minimum energy required to add one fermion to the unpolarized unitary Fermi gas. This quantity plays an important role in characterizing the superfluidity of the system.

In the unitary regime the interaction is strongly attractive and non-perturbative calculations are required. Since the details of the microscopic interaction are unimportant, such a system can be effectively described by four-Fermi contact operators with two-component fermions in the Lagrangian. Based on this theory, we recently developed a highly improved lattice calculation method to explore a large number of strongly interacting non-relativistic fermions \cite{endres}. We briefly review this method in Sec. 2. There has also been a substantial number of numerical studies of the unitary fermions from the microscopic theory using the Quantum Monte Carlo \cite{qmc1, qmc2, quasi, qmc3} and various other Monte Carlo techniques \cite{dean, abe}.

The underlying idea of the lattice theory is that the correlation function decays with the ground state energy in the long imaginary time limit. In principle, this is true as long as the interpolating operator has some overlap with the ground state and the error propagation is moderate with time. However, practically it is very difficult to find the ground state in simulating many interacting particles. Because the overlap between the source and the ground state decreases exponentially with respect to the number of particles ($N$) and the accessible time range is relatively small for large $N$, such as our case, as well. One way of overcoming these issues is by constructing improved sources (or sinks), which have large overlap with the ground state and result in earlier plateaus in effective mass plots. In Sec. 3, we discuss our choice of sources in detail and demonstrate the improvement that can be achieved with an example. In Sec. 4, with these improvements we calculate the ground state energy of up to $38$ unitary fermions in a finite box and report our preliminary results for $\xi$ and $\Delta$. In our companion proceedings, we present a similar calculation for up to $20$ unitary fermions in a harmonic trap \cite{nicholson}.

\section{Lattice construction for fermions at unitarity}
In this section, we briefly review the lattice construction for unitary fermions which is discussed in Ref. \cite{endres} in detail. Starting with the lattice action for interacting non-relativistic two-component fermions which is first introduced in Ref. \cite{chen}, we construct the single fermion propagator
\beq
K^{-1}(\tau;0)=D^{-1}(1-\phi(\tau-1)\sqrt{C})D^{-1}\cdots D^{-1}(1-\phi(0)\sqrt{C})D^{-1}, \label{single_prop}
\eeq
where $D^{-1}=1-\nabla^2/2M$, $\phi(\tau)$ is a $Z_2$ auxiliary field on each time-link produced stochastically, and $C$ is a generalized version of four-Fermi coupling involving an effective derivative operators with appropriate coefficients. With open boundary conditions in the time direction and time-like links of $\phi$s, one can show that the fermion determinant is independent of the auxiliary field and the full numerical simulation is quenched. By averaging out the $\phi$s, one can achieve a four-Fermi interaction. The $N$-fermion propagator can be constructed by taking direct products of Eq. (~\ref{single_prop}), anti-symmetrizing, and then ensemble averaging.

In order to minimize the lattice artifacts, we consider a couple of lattice improvements. First, we use $D$ in momentum space by
\begin{equation}
D({\bf p})=\left\{\begin{array}{c}
e^{p^2/(2M)}\delta_{\bf p,p'}, ~~~~~p<\Lambda \\
\infty ~~~~~~~~~~~~~~~~~, ~~~~~~p\geq \Lambda
\end{array}
\right.,
\end{equation}
where $\Lambda=\pi$ is a hard momentum cutoff and $p=|{\bf p}|$. With this definition of $D$, one can eliminate the discretization errors appearing in free fermion propagator. Second, we minimize the discretization errors associated with the contact interaction on the lattice when we tune the coupling constants to describe the unitary limit. The key idea of our tuning procedure is the following : we tune the coupling constants by matching the low-lying energy eigenvalues calculated from the two-particle transfer matrix $\mathcal{T}$ in our lattice theory to those of a finite volume and continuum theory for $p\cot\delta_0=0$ using Luscher's formula. In principle, with these tuning methods we can eliminate both discretization errors and finite volume effects. Actually, in the unitary limit these two errors are equivalent since the only relevant dimensionless quantity is $b/L$, where $b$ is the lattice spacing and $L$ is the size of the lattice.

\section{$N$-particle correlation function and two-particle source}
Since we are working in the canonical ensemble, we consider a two-component system of $N_\uparrow$ spin up and $N_\downarrow$ spin down fermions. To satisfy Fermi-Dirac statistics, fermions of the same species have different quantum numbers. For the purpose of measuring correlators, we take the lowest momentum eigenstates of an uncorrelated single free fermion as sources, such as ${\bf p}_1=(0,0,0)2\pi/L,~{\bf p}_2=(1,0,0)2\pi/L,~\cdots$ in the Cartesian basis of a cubic lattice with size $L$.

For the unpolarized fermi gas, the $N$-body correlation function $C_N(\tau)$ ($N_\uparrow=N_\downarrow=N/2$) can be obtained from a Slater determinant of the $N/2 \times N/2$ matrix as
\beq
C_N(\tau)=\langle \textrm{det}({\bf S}(\tau))\rangle, ~~~~~~~~S_{ij}(\tau)=\langle \psi_{sink} |K^{-1}(\tau;0)\otimes K^{-1}(\tau;0) | {\bf p}_i {\bf p}_j\rangle,
\eeq
where $\psi_{sink}$ is an interpolating field of a spin up and a spin down fermion for the sink, and $i,j=1,2,\cdots N/2$. 
In the case of $N_\uparrow=N_\downarrow+1$, we replace the $k$th row by $\langle {\bf p}_k |K^{-1}(\tau;0)| {\bf p}_j\rangle$ with $j=1,2,\cdots,N/2$.

The sinks may be chosen as the lowest momentum eigenstates of an uncorrelated single free fermion as well, which corresponds to the ground state of the non-interacting fermions \cite{dean}. However, the two-particle $S$-wave solutions to the Schrodinger equations for unitary and non-interacting fermions behave like $1/r$ and $1$, respectively, where $r$ is the relative distance between the two fermions. In order to have better overlap of the interpolating field with the ground state of $N$ unitary fermions in a finite box, we choose the sink for one spin-up and one spin-down fermion as a two-particle state given by
\beq
\psi_{sink}({\bf p})=
\left\{
\begin{array}{c}
\frac{e^{-b p}}{p^2}, ~~~~~~p\neq 0 \\
\psi_0, ~~~~~~~~~p= 0
\end{array}
\right.
\label{pair}
\eeq
where ${\bf p}$ is the momentum of each fermion in the center of momentum frame, $p=|{\bf p}|$, and $b$ and $\psi_0$ are tunable parameters. Ignoring the $p\neq 0$ component, $\psi_{sink}({\bf p})$ with $b=0$ corresponds to the wave function of the unitary fermions in an infinite volume and continuum theory. The result is not sensitive to $\psi_0$ and we consider $\psi_0=100$ in this work.

\begin{figure}
\bmp[t]{.48\linewidth}
\centering
\includegraphics[width=1.0\textwidth]{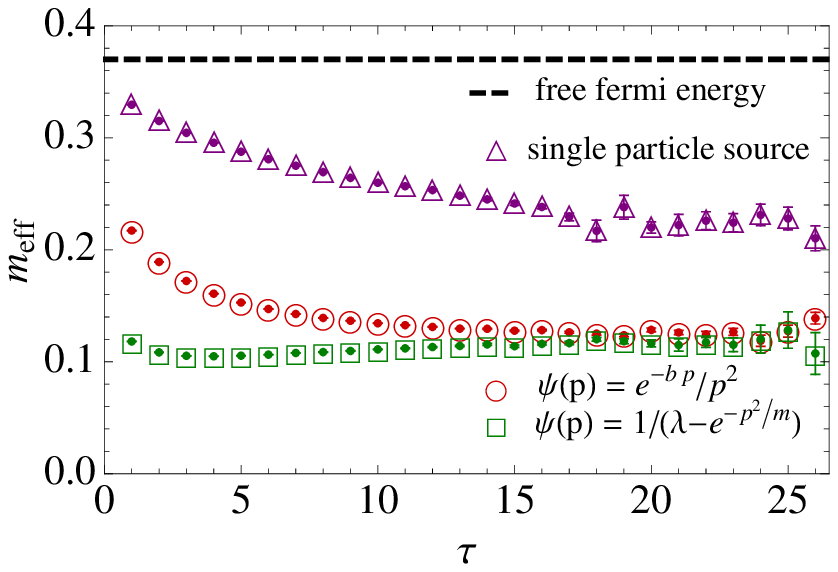}
\caption{Effective mass plots for $L=8$, $M=5$, and $N=8$ ($N_\uparrow=N_\downarrow=4$). The purple triangles are obtained using the lowest momentum eigenstates for a uncorrelated single free fermion while red circles and green squares are obtained using correlated two-particle states for spin up and down as sinks. The black dashed line correspond to the ground state energy for non-interacting fermions. $b=0.5$ and $\lambda=1.01$ are used for red and green, respectively.} \label{fig1}
\emp
\hskip .2in
\bmp[t]{.48\linewidth}
\includegraphics[width=1.0\textwidth]{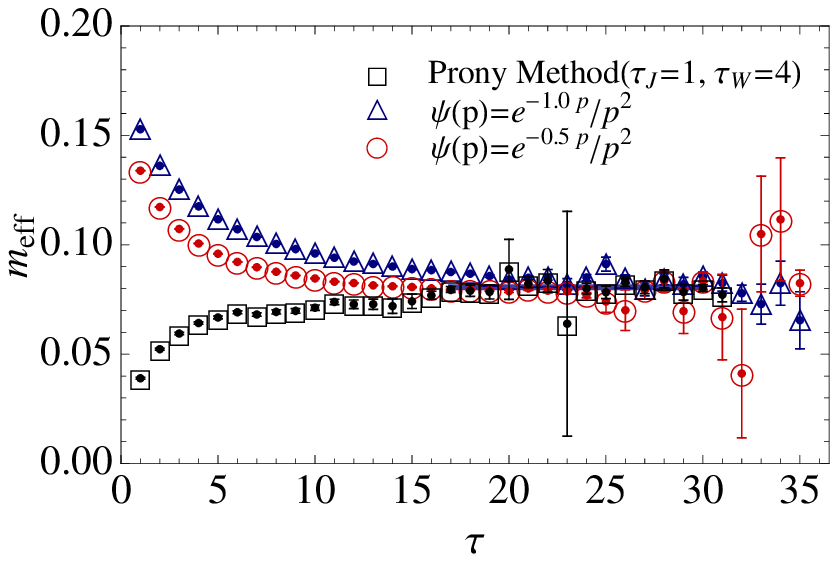}
\caption{Effective mass plots for $L=12$, $M=5$, and $N=10$ ($N_\uparrow=N_\downarrow=5$). The red circles and blue triangles are results using the sinks in Eq. (3.2) with $b=0.5$ and $b=1.0$, respectively. The black squares are obtained using the Multi-Correlation function Prony Method with the data from these two sinks, where the parameter $\tau_J$ and $\tau_W$ are introduced to improve stability \cite{beane}. } \label{fig2}
\emp
\end{figure}

As an example of the wavefunction overlap problem in many particle calculations and the importance of using a proper choice for the interpolating operator, we plot the effective masses with three different sinks for $L=8$, $M=5$, and $N=8$ in \reffig{fig1}. Around $\tau=18-25$, the effective masses with two different correlated two-particle sinks, red (corresponding to the operator in Eq. (3.2)) and green (corresponding to the eigenstate of the two particle transfer matrix with a tunable parameter $\lambda$), have clean plateaus with consistent fit values while the effective mass with the uncorrelated single particle sink (purple) is much higher and has no clear plateau. This implies that the purple doesn't reach to the ground state, yet. For $\tau>26$ all of the effective mass plots become very noisy. This result shows that it is very difficult or even impossible to see the unitary fermion ground state using sinks constructed from single particle momentum eigenstates. 

\section{Preliminary Results}

In this work, we calculate the ground state energy of $N\leq 38$ unitary fermions and extract preliminary results for the Bertsch parameter and pairing gap. As discussed in the previous section, the plateaus in the effective mass plots are much clearer when the sinks are constructed from correlated two-particle states. Since we have different sinks for the calculation, we can obtain slightly longer and better plateaus using the Multi-Correlation Prony Method \cite{beane} with two different sinks (\reffig{fig2}). After blocking the data, we perform a Jackknife resampling procedure and then a $\chi^2$ correlated fit of the plateau obtained using the Prony Method. The statistical error is obtained by varying the fit parameter to give $\chi^2=\chi_{min}^2+1$ while the fitting systematic error is obtained by varying the end points of the fitting range by $\pm 2, \pm 1, 0$. The quoted errors represent only the combination of the fitting statistical and systematic errors in quadrature and does not include the systematic errors from finite volume effects or discretization errors.

\begin{figure}
\bmp[t]{.48\linewidth}
\centering
\includegraphics[width=1.0\textwidth]{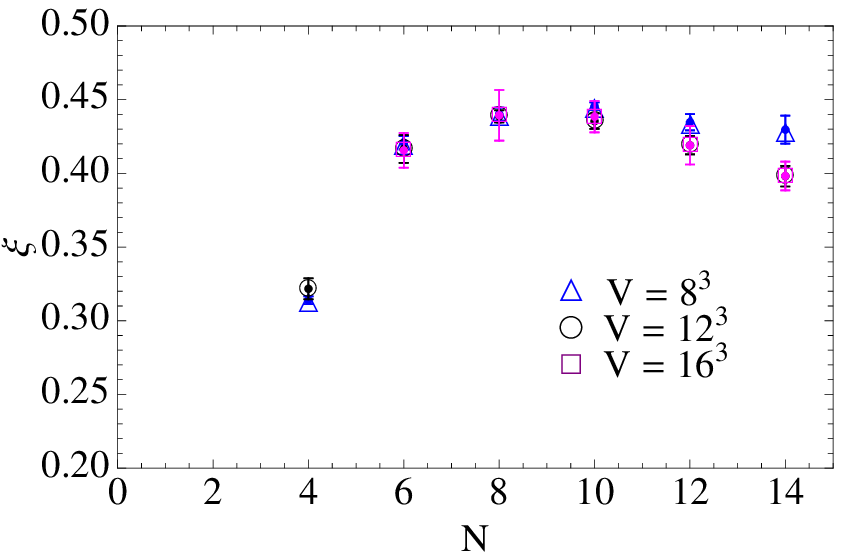}
\caption{(Preliminary) Bertsch parameter as a function of $N$ for $V=8^3$ (blue), $12^3$ (black), and $16^3$ (purple).} \label{fig3}
\emp
\hskip .2in
\bmp[t]{.48\linewidth}
\centering
\includegraphics[width=1.0\textwidth]{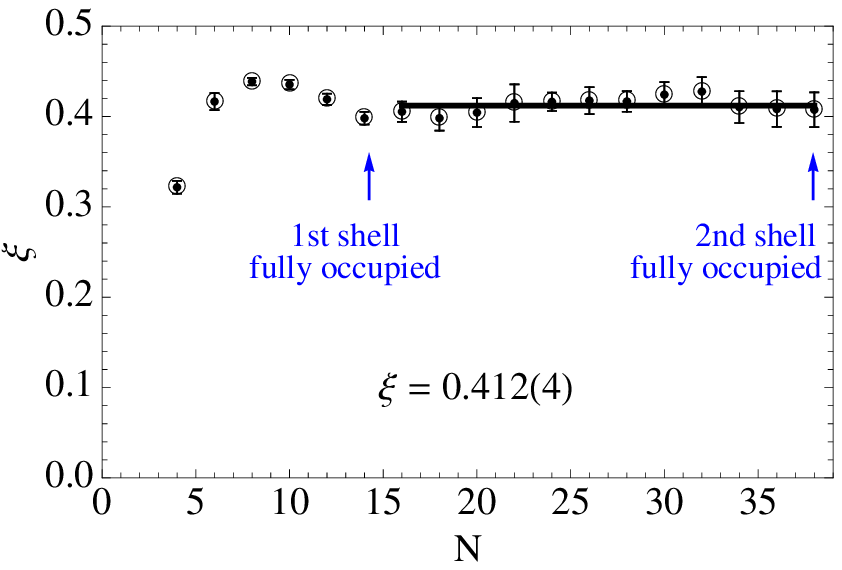}
\caption{(Preliminary) Bertsch parameter as a function of $N$ for $V=12^3$. The solid line represents the fitting range and fit result for $\xi$.} \label{fig4}
\emp
\end{figure}

As discussed in Sec. 2, in principle we expect that there should be no volume dependence in the Bertsch parameter. However, since we only tune a finite number of the coupling constants, the deviation of $p\cot\delta$ from zero gets larger with $p$, resulting in the presence of discretization errors for small $V$ and large $N$. In \reffig{fig3}, we calculate $\xi$ for $N\leq 14$ unitary fermions in a periodic box with $V=8^3$, $12^3$, and $16^3$. For $N\geq 14$, $\xi$ with $V=8^3$ is slightly greater than that of the other two volumes. This result shows that $V=8^3$ is too small to calculate the correct $\xi$ for $N\geq 14$.

The results for the Bertsch parameter for $N\leq 38$ and $V=12^3$ are shown in \reffig{fig4}\footnote[1]{We also did the simulation for $L=16$, but for the given computer resources we did not have sufficient statistics to calculate the Bertsch parameter for the second shell.}. We find that $\xi$ is not independent of $N$ for the first shell while we don't see the $N$ dependence in the second shell with given statistics. A fit to the second shell gives $\xi=0.412(4)$ (preliminary). 
The value of $\xi$ has been obtained from various theoretical calculations and ultra-cold atomic experiments. For example, the recent QMC calculation gives $\xi= 0.40(1)$ \cite{gezerlis} while one of the recent experimental measurements shows $\xi=0.39(2)$ and $\xi=0.41(2)$ using two different extraction methods \cite{Luo}. 

\begin{figure}
\bmp[t]{.48\linewidth}
\centering
\includegraphics[width=1.0\textwidth]{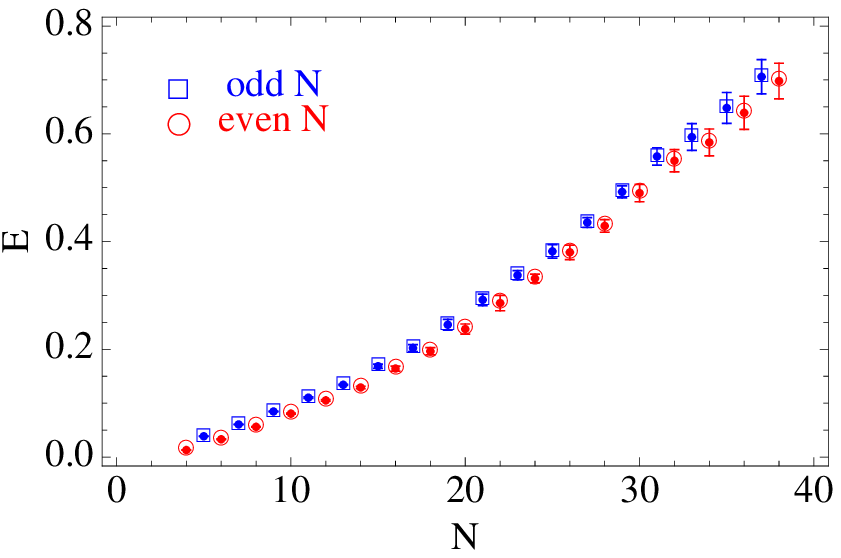}
\caption{(Preliminary) Ground state energies for $N\leq 38$. The red circle and blue square represent the ground state energies for unpolarized and slightly polarized unitary fermions.} \label{fig5}
\emp
\hskip .2in
\bmp[t]{.48\linewidth}
\centering
\includegraphics[width=1.0\textwidth]{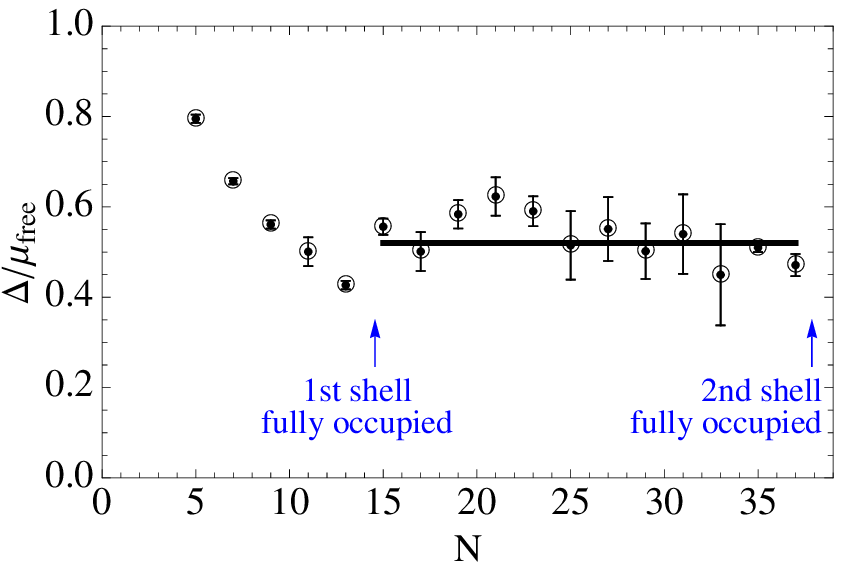}
\caption{(Preliminary) Pairing gap in unit of $\mu_{free}$ for $N\leq 37$, where $\mu_{free}$ is the free Fermi energy. The solid line represents the fitting range for the extracted value of $\Delta$.} \label{fig6}
\emp
\end{figure}

We also perform the simulation for the slightly polarized unitary fermi gas, $N_\uparrow=N_\downarrow+1$. As discussed in Ref. \cite{quasi}, the ground state energy for odd $N$ depends on the momentum of the unpaired spin up fermion. For $N\leq 37$, the minimum energies are obtained by placing the unpaired fermion on the first momentum shell, and used for \reffig{fig5} and the gap calculation. In \reffig{fig5}, we plot the ground state energy of both the unpolarized and the slightly polarized unitary fermi gas. The staggering between even and odd indicates a non-zero pairing gap. Using this data, we calculate the pairing gap as
\beq
\Delta = E(N_\uparrow,N_\downarrow)-\frac{E(N_\uparrow,N_\downarrow+1)+E(N_\uparrow-1,N_\downarrow)}{2},
\eeq
and the results are shown in \reffig{fig6}. By fitting the data in the second momentum shell, we obtain $\Delta=0.52(1)$. The most recent QMC calculation \cite{quasi} and ultra-cold atomic experiment \cite{kettler} reported $\Delta=0.50(3)$ and $\Delta=0.43(3)$, respectively. 

\section{Conclusion}
In these proceedings, we explored up to $38$ unitary fermions in a finite box, and extracted preliminary values for the Bertsch parameter and pairing gap. The absence of volume dependence, a consequence of the tuning method for the four-Fermi contact interaction, allows us to study the infinite volume, continuum properties of unitary fermions with a moderately sized lattice. Our analysis also shows that a careful choice of the interpolating operators, which are designed to have large overlap with the state of interest, plays a crucial role in simulating the interacting many-particle system on the lattice. In the future, we hope to extend our simulations to larger numbers of unitary fermions using a recently developed tuning procedure\footnote[2]{We have recently developed a new tuning method which has Hermitian, Galilean invariant and analytic interaction.} and more careful study of the correlated two-particle sinks.

\section{Acknowledgement}
This work was supported by U. S. Department of Energy grants DE-FG02-00ER41132 (to D. B. K., J-W. L. and A. N. N.) and DE-FG02-92ER40699 (to M. G. E.). This research utilized resources at the New York Center for Computational Sciences at Stony Brook University/Brookhaven National Laboratory which is supported by the U.S. Department of Energy under Contract No. DE-AC02-98CH10886 and by the State of New York.


\begin{thebibliography}{99}
\bibitem{endres} M. G. Endres, D. B. Kaplan, J.-W. Lee, and A. N. Nicholson, 
\pos{PoS LATTICE2010, 182 (2010)}.

\bibitem{qmc1} J. Carlson, S.-Y. Chang, V. R. Pandharipande, and K. E. Schmidt, 
Phys. Rev. Lett. {\bf 91}, 050401 (2003).

\bibitem{qmc2} G. E. Astrakharchik, J. Boronat, J. Casulleras, and S. Giorgini, 
Phys. Rev. Lett. {\bf 93}, 200404 (2004), cond-mat/0406113.

\bibitem{quasi} J. Carlson and S. Reddy, 
Phys. Rev. Lett. {\bf 95}, 060401 (2005), cond-mat/0503256.

\bibitem{qmc3} A. Bulgac, J. E. Drut, and P. Magierski, 
Phys. Rev. A {\bf 78}, 023625 (2008), arXiv:0803.3238.

\bibitem{dean} Dean Lee, 
Phys. Rev. B {\bf 73}, 115112 (2006), cond-mat/0511332.

\bibitem{abe} T. Abe and R. Seki, 
Phys. Rev. C {\bf 79}, 054003 (2009), nucl-th/0708.2524.

\bibitem{nicholson} A. N. Nicholson, M. G. Endres, D. B. Kaplan, and J.-W. Lee, 
\pos{PoS LATTICE2010, 206 (2010)}.

\bibitem{chen} J.-W. Chen and D. B. Kaplan, 
Phys. Rev. Lett. {\bf 92}, 257002 (2004), hep-lat/0308016.

\bibitem{beane} Silas R. Beane \emph{et al}, 
Phys. Rev. D {\bf 79}, 114502 (2009), hep-lat/0903.2990.

\bibitem{gezerlis} A. Gezerlis and J. Carlson, 
Phys. Rev. C {\bf 77}, 032801 (2008), nucl-th/0711.3006.

\bibitem{Luo} L. Luo and J. E. Thomas,  
Journal of Low Temperature Physics {\bf 154}, 1 (2009), 0811.1159.

\bibitem{kettler} A. Schirotzek, Y. Shin, C. Schunck, and W. Ketterle, 
Phys. Rev. Lett. {\bf 101}, 140403 (2008).

\end{thebibliography}
\end{document}